\documentclass[preprint,12pt]{elsarticle}
\usepackage{geometry}  
\usepackage{graphicx}  
					   
\usepackage{tikz}
\usepackage[compat=1.1.0]{tikz-feynman}

\newcommand{\be}{\begin{equation}}
\newcommand{\ee}{\end{equation}}
\newcommand{\bea}{\begin{eqnarray}}
\newcommand{\eea}{\end{eqnarray}}
\newcommand{\bean}{\begin{eqnarray*}}
\newcommand{\eean}{\end{eqnarray*}}

\def\bra#1{\left\langle #1\right|}
\def\ket#1{\left| #1\right\rangle}

\begin{document}

\title{The role of long distance contribution to the $B\to K^{(\ast)} \ell^+ \ell^-$ in the Standard Model}

\author[CNR]{Massimo Ladisa}
\address[CNR]{CNR, Istituto per le Applicazioni del Calcolo ``M. Picone", Via Amendola 122-D, I-70126 Bari, Italy}
\author[Unina,INFN]{Pietro Santorelli}
\address[Unina]{Dipartimento di Fisica ``E. Pancini", Universita' di Napoli Federico II - 
Complesso Universitario di Monte S. Angelo Edificio 6, via Cintia, 80126 Napoli, Italy}
\address[INFN]{INFN sezione di Napoli -  Complesso Universitario di Monte S. Angelo Edificio 6, via Cintia, 80126 Napoli, Italy}

\begin{abstract}
We investigate rare semileptonic $B \to K^\ast\ell^+\ell^-$ by looking at the 
long distance contributions. 
Our analysis is limited to the very small values of physical accessible range of invariant
mass of the leptonic 
couple $q^2$. We show that the light quarks loop has to be accounted for, 
along with the charming penguin contribution, in order to accurately compute the 
$q^2$-spectrum in the Standard Model. 
Such a long distance contribution may also play a role in the
analysis of the lepton flavour universality violation in this process.
\end{abstract}

\maketitle

\vspace{1cm}

\noindent
In the Standard Model the Flavour-Changing Neutral Current (FCNC) processes are sensitive probes of New Physics (NP) because they arise at loop level and are further suppressed by GIM mechanism. 
A reliable calculation of the process in the framework of the Standard Model is the first step to highlight effects of NP.
The next one consists in comparing the hopefully precise measurements with the theoretical calculations. 
For a thorough review on the subject we address the reader to the recent paper \cite{DAlise:2022ypp} and the exhaustive bibliography therein.
Hereinafter we will discuss one possible Long Distance (LD) contribution to the exclusive process $B\to K^\ast \ell^+ \ell^-$ which in principle is Cabibbo--Kobayashi--Maskawa (CKM) suppressed. We start discussing the $e^2$ corrections to the 
effective Hamiltonian responsible, in the Standard Model, of the $B\to K^{(\ast)} \ell^+ \ell^-$ decay, then we evaluate them. 

\par 
The effective Hamiltonian for $\Delta B =-\Delta S = 1$  in the Standard Model responsible of the rare transition $b \to s \ell^+ \ell^-$ can be written in terms of a set of local operators \cite{GSW}: 
\begin{equation}
{\cal H}_W = \frac{4 G_F}{\sqrt{2}} V_{tb} V_{ts}^\ast \sum_{i=1}^{10} C_i(\mu)
O_i(\mu) = {\cal H}^{had} +  {\cal H}^{sl} + {\cal H}^{\gamma}\ ,
\label{hamil} 
\end{equation}
\noindent 
where $G_F$ is the Fermi constant and $V_{ij}$ are elements of the 
CKM mixing matrix.
The operators $O_i$, written in terms  of quark, photon and gluon fields, and can be found
for example in Ref. \cite{Chetyrkin:1996vx} and the ${\cal H}^{had}$ contains the operators $O_i$ with 
$i = 1,\ldots ,6$,  ${\cal H}^{sl}$ contains the operators $O_9$ and $O_{10}$
whereas ${\cal H}^{\gamma}$ contains $O_7$.\footnote{$O_8$ is the analogous $O_7$ for the gluon fields.} 
In our calculation the main role is played by the operators $O_i$ with $i \in \{1,2,7,9,10\}$ 
which we report here for convenience 
\begin{eqnarray}
O_1&=&({\bar s}_{L \alpha} \gamma^\mu b_{L \alpha})
      ({\bar c}_{L \beta} \gamma_\mu c_{L \beta}) \ ,\nonumber \\
O_2&=&({\bar s}_{L \alpha} \gamma^\mu b_{L \beta})
      ({\bar c}_{L \beta} \gamma_\mu c_{L \alpha}) \ ,\nonumber \\
O_7&=&\frac{e}{16 \pi^2} m_b ({\bar s}_{L \alpha} \sigma^{\mu \nu} 
     b_{R \alpha}) F_{\mu \nu} \ ,\nonumber \\
O_9&=&\frac{e^2}{16 \pi^2}  ({\bar s}_{L \alpha} \gamma^\mu 
     b_{L \alpha}) \; {\bar \ell} \gamma_\mu \ell\ , \nonumber \\
O_{10}&=& \frac{e^2}{16 \pi^2}  ({\bar s}_{L \alpha} \gamma^\mu 
     b_{L \alpha}) \; {\bar \ell} \gamma_\mu \gamma_5 \ell \ .
\label{operators} 
\end{eqnarray}
\noindent 
The greek letters are  colour indices and, as usual, 
$ b_{R/L}=\left(\frac{1 \pm \gamma_5}{2}\right)b$, and
$\sigma^{\mu \nu}=\frac{i}{2}[\gamma^\mu,\gamma^\nu]$. $F_{\mu \nu}$ denotes 
the electromagnetic field strength tensor and $e$ is the electromagnetic charge. \par
From here on we shall focus on the $B\to K^{\ast}\,\ell^+\ell^-$ process. In Ref \cite{noibis} we will systematically analyse the processes containing the $K$ and the $K^*$. \par
There are two class of contributions to the $B\to K^{\ast}\, \ell^+\, \ell^-$, 
the first one comes from the  semileptonic part of the effective Hamiltonian, i.e. the
operators $O_9$ and $O_{10}$. In this case the amplitude factors out
\begin{eqnarray}
&& \hspace{-2truecm} A_{\rm SD}(B\to K^{\ast}\,\ell^+\ell^-)   =  
\bra{K^{\ast}\,\ell^+\ell^-}{\cal H}^{sl} \ket{B} = \nonumber\\
&& \frac{4 G_F}{\sqrt{2}} V_{tb} V_{ts}^*\frac{e^2}{16\pi^2}
 \sum_{i=9,10}C_i \bra{\ell^+\ell^-} {\bar \ell} \Gamma_\mu^i \ell \ket{0}
\bra{K^{(*)}} ({\bar s}_{L \alpha} \gamma^\mu b_{L \alpha}) \ket{B}\ ,
\label{e:SD}
\end{eqnarray}
and it can be written in terms of form factors (e.g. those in Ref. \cite{Colangelo:1995jv}, 
our choice hereafter). This is called the short distance (SD) part of the total amplitude 
of the process: the hadronic contribution is incorporated in the form factors while the 
perturbative corrections in the Wilson coefficients (for them we use the same values in \cite{Colangelo:1995jv}).
\par
\begin{figure}[h]
\begin{center}
\begin{tikzpicture}
\small
\begin{feynman}
  \vertex (a1) {\(b\)};
  \vertex[right=1cm of a1, square dot] (a2) {\(\)};
  \vertex[right=1.5cm of a2] (a3);
  \vertex[above=0.5cm of a3] (a4);
  \vertex[below=2cm of a4] (a5);
  \vertex[right=1.5cm of a2] (a6);
  \vertex[above=0.875cm of a6] (a7); 
  \vertex[right=1cm of a7] (a8) {\(s\)};
  \vertex[below=0.375cm of a8] (b1) {\(\overline u\)};
  \vertex[left=1cm of b1] (b2);
  \vertex[below=2.5cm of b2, dot] (b3) {\(\)};
  \vertex[below=0.375cm of a1] (c1) {\(\overline u\)};
  \vertex[right=1cm of c1] (c2);
  \vertex[below=2.25cm of b1, dot] (b4) {\(\)};
  \vertex[right=1.75cm of a5] (lm) {\(\ell^-\)};
  \vertex[below=0.625cm of lm] (lp) {\(\ell^+\)};
  \diagram* {
  {[edges=fermion] (a1) -- (a2)},
  {[edges=anti fermion] (a2) -- [edge label'={\(\overline u\)}] (a4) -- (a5) -- [edge label'={\(u\)}](a2)},    
  {[edges=fermion] (a2) --  [edge label={}] (a7) -- (a8)}, 
  {[edges=fermion] (b1) -- (b2) -- [edge label={}] (b3)}, 
  {[edges=anti fermion] (c1) -- (c2) --[edge label'={}]  (b3)}, 
  {[edges=photon] (b3) -- [edge label={\(\gamma^\star\)}] (b4)},
  {[edges=fermion] (b4) -- (lm)},
  {[edges=anti fermion] (b4) -- (lp)}
  };
      \draw [decoration={brace}, decorate] (c1.south west) -- (a1.north west) node [pos=0.5, left] {\(B^-\)};
      \draw [decoration={brace}, decorate] (a8.north east) -- (b1.south east) node [pos=0.5, right] {\(K^{\star -}\)};
\end{feynman}
\end{tikzpicture}
\caption{\small The Feynman graph obtained by doing the $T$-product in Eq. (\ref{e:Hu})}
\label{f:T-prod}
\end{center}
\end{figure}
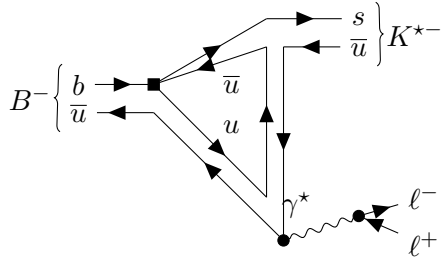
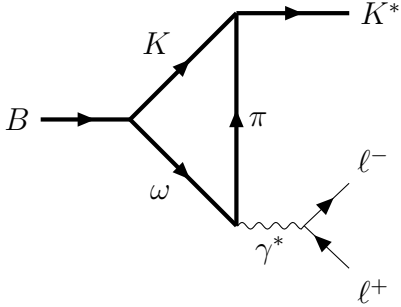
\begin{figure}
\begin{center}
\begin{tikzpicture}
\begin{feynman}
\vertex (a1) {\(B\)};
\vertex [right=of a1 , square dot] (a2);
\vertex [above right = 2cm of a2] (a3);
\vertex [below right = 2cm of a2] (a4);
\vertex [right=of a3] (b1) {\(K^\ast\)};
\vertex [right= 0.9 cm of a4] (b2);
\vertex [above right = 0.8 cm of b2] (c1){\(\ell^-\)};
\vertex [below right =0.8 cm of b2] (c2){\(\ell^+\)};
\diagram* {
(a1) -- [fermion, ultra thick ] (a2) -- [fermion, ultra thick, edge label=\(K\) ] (a3);
(a2) -- [fermion, ultra thick, edge label'=\(\omega\) ] (a4) -- [fermion, ultra thick, edge label'=\(\pi\)] (a3);
(a3) -- [fermion, ultra thick ] (b1);
(a4) -- [boson, edge label'=\(\gamma^\ast\)] (b2);
(b2) -- [fermion]  (c1);
(b2) -- [anti  fermion] (c2);
};
\end{feynman}
\end{tikzpicture}

\caption{\small Hadronic representation of one of the contributions coming from the graph in Fig. \ref{f:T-prod} .}
\label{f:T-prodMes}
\end{center}
\end{figure}

Moreover, at the same order in $e$, i.e. $e^2$, the amplitude contains the contribution of the  hadronic effective Hamiltonian multiplied by the QED interaction twice. 
An amplitude different from zero is obtained when the former interaction produces the leptonic 
pair in the final state and the latter one factors an hadronic current out:
\begin{eqnarray}
&&\hspace{-0.8truecm}A_{LD} (B\to K^{\ast}\,\ell^+\ell^-)   = \nonumber\\
&&\hspace{-0.8truecm}e^2 \bra{K^{\ast}\,\ell^+\ell^-}T \int A^\mu(x)  \bar \ell (x) \gamma_\mu\ell (x) dx \int dy \left[A^\nu(y)  j_\nu^{e.m.}(y) \right] {\cal H}^{had}(0) \ket{B} = \nonumber\\
&&\hspace{-0.8truecm}-\frac{i e^2}{q^2}\int d^4x e^{-iqx}\bra{\ell^+\ell^-}\bar\ell(x)\gamma_\mu\ell(x) \ket{0}  
\int d^4y e^{i qy} \bra{K^*(p^\prime)} T j^{\nu}_{e.m.}(y) {\cal H}^{had}(0) \}\ket{B(p)} \nonumber\\
&&\hspace{-0.8truecm}\equiv L_\mu {\cal H}^\mu(p,p^\prime)\,.
\label{e:nonlocal}
\end{eqnarray}
${\cal H}^\mu(p,p^\prime)$ is essentially a non-local term and we call it LD 
contribution to the
decay process although, as discussed recently for example in \cite{Gubernari:2020eft}, the
hadronic matrix
element ${\cal H}(p,p^\prime)$ contains a factorizable part.  We shall clarify this point.
In (\ref{e:nonlocal})  $q^2=(p-p^\prime)^2$ and $ j_\mu^{e.m.}= \sum_q Q_q\,\bar q \gamma_\mu q$. 
By considering the CKM matrix elements
and the strength of the Wilson coefficients we can conclude that the leading contribution to 
$\cal H$ will come from the operators $O_1$ and
$O_2$ in the effective Hamiltonian  proportional to $V_{cb} V_{ cs}^*\approx V_{tb} V_{ts}^*$ 
and so the T-product is different from zero if and only if  
$\displaystyle j_\mu^{e.m.}= Q_c\,\bar c \gamma_\mu c$: 
the  contribution of these terms is commonly
called the {\it charm-loop effect}. In other words:
\begin{eqnarray}
{\cal H}^\mu =Q_c  \int d^4y e^{i qy} \bra{K^*(p^\prime)} {\rm T}\, \bar c(y) \gamma^\mu c(y) \left (C_1 O_1(0) + C_2 O_2(0) \right )\ket{B(p)}\ .
\end{eqnarray}
The analysis, in QCD -factorization, of the non-local term in the previous equation was done in \cite{Beneke:2001at} where the LD contribution results to be essentially proportional to the factorizable part, i.e. our $A_{SD}$. 
A systematic study of ${\cal H}^\mu$ can be read in \cite{Khodjamirian:2010vf} where the authors show how to
generalize the approach in \cite{Beneke:2001at} to the low-$q^2$ region. The state-of-the-art of these calculations can be found in Ref.  \cite{Gubernari:2022hxn}. Due to the difficulties to reliable estimate the LD contribution, a different approach relies on the use of data-driven methods 
to account for the theoretical uncertainties and to
quantify possible deviations from the Standard Model \cite{Ciuchini:2021smi}. All these papers are devoted to the charm loop contribution. Hereinafter we shall study a contribution which is CKM suppressed although it gives, at very small $q^2$, a contribution comparable to the short distance one. At this aim, the four quarks operators are
\begin{eqnarray}
O_1^{(u)}&=&({\bar s}_{L \alpha} \gamma^\mu b_{L \alpha})
      ({\bar u}_{L \beta} \gamma_\mu u_{L \beta}) \ , \nonumber \\
O_2^{(u)}&=&({\bar s}_{L \alpha} \gamma^\mu b_{L \beta})
      ({\bar u}_{L \beta} \gamma_\mu u_{L \alpha}) \ , 
\label{operatorsCKM} 
\end{eqnarray}
and ${\cal H}^\mu$ is different from zero for the $j_\mu^{e.m.}= Q_u\,\bar u \gamma_\mu u$:
\begin{eqnarray}
{\cal H}^{(u)\mu} =Q_u  \int d^4y e^{i qy} \bra{K^*(p^\prime)} {\rm T}\, \bar u(y) \gamma^\mu u(y) 
\left (C_1 O_1^{(u)}(0) + C_2 O_2^{(u)}(0) \right )\ket{B(p)} \,;
\label{e:Hu}
\end{eqnarray}
in Fig. \ref{f:T-prod} one can find the Feynman graph of the T-product in Eq. (\ref{e:Hu}) while in Fig. \ref{f:T-prodMes} a possible 
mesonic graph, the one we shall consider hereafter\footnote{The graph 
obtained by interchanging $\omega$ and $\pi$ mesons is suppressed  
by the off-shellness of both the $\omega$ meson in the t channel and 
the $K$-$K^\ast$-$\omega$ coupling.}.  We stress that this is just one of the possible hadronic terms representing 
${\cal H}^{(u)\mu}$ and so  our calculation is just an estimation of one non-factorizable, LD contribution to the one in Eq.(\ref{e:SD}).
Unless expressely stated in a theorem, we rule out the hypothesis that summing up several of such contributions would conjure in a global cancellation.

\noindent
The calculation of the triangle graph in Fig.  \ref{f:T-prodMes}  is straightforward. 
We consider the weak transition $B^-\to K^-\omega $ followed by the electromagnetic one 
$\omega\to \pi^0 \gamma$. The meson-loop is closed by the $K^\ast$-$K$-$\pi^0$ strong vertex, $g_{KK^\star \pi}$, computed by the $K^\ast \to K \pi^0$ decay.
The analysis of the $B^-\to K^-\omega $ transition was done, for example, in Ref. \cite{Isola:2003fh} where the contribution of the charming 
penguins was accounted for to improve the factorization approximation prediction. 
In particular, the branching ratio of 
$B^-\to K^-\omega$ is enhanced by the charming penguin contribution of about one order of magnitude; our prediction \cite{Isola:2003fh},  
$Br(B^-\to K^-\omega) = 6.19 \,\times\, 10^{-6}$, is in excellent agreement with the PDG average  
$Br(B^-\to K^-\omega) =  (6.5 \,\pm\, 0.4)\ \times\, 10^{-6}$  \cite{Workman:2022ynf}. 
In Ref. \cite{Isola:2003fh} the so called charming penguins, discussed for the first time in \cite{Ciuchini:1997hb}, are evaluated by 
considering  the charm rescattering into the charmless two body final state, i.e., for example, 
$B \to D\, D _s\to K\,\omega$ (cfr also  \cite{Isola:2001ar,Isola:2001bn,Ladisa:2004bp,Deandrea:2005ii}).\par
\begin{figure}[!htp]
    \centering
    \includegraphics[width=0.8\textwidth]{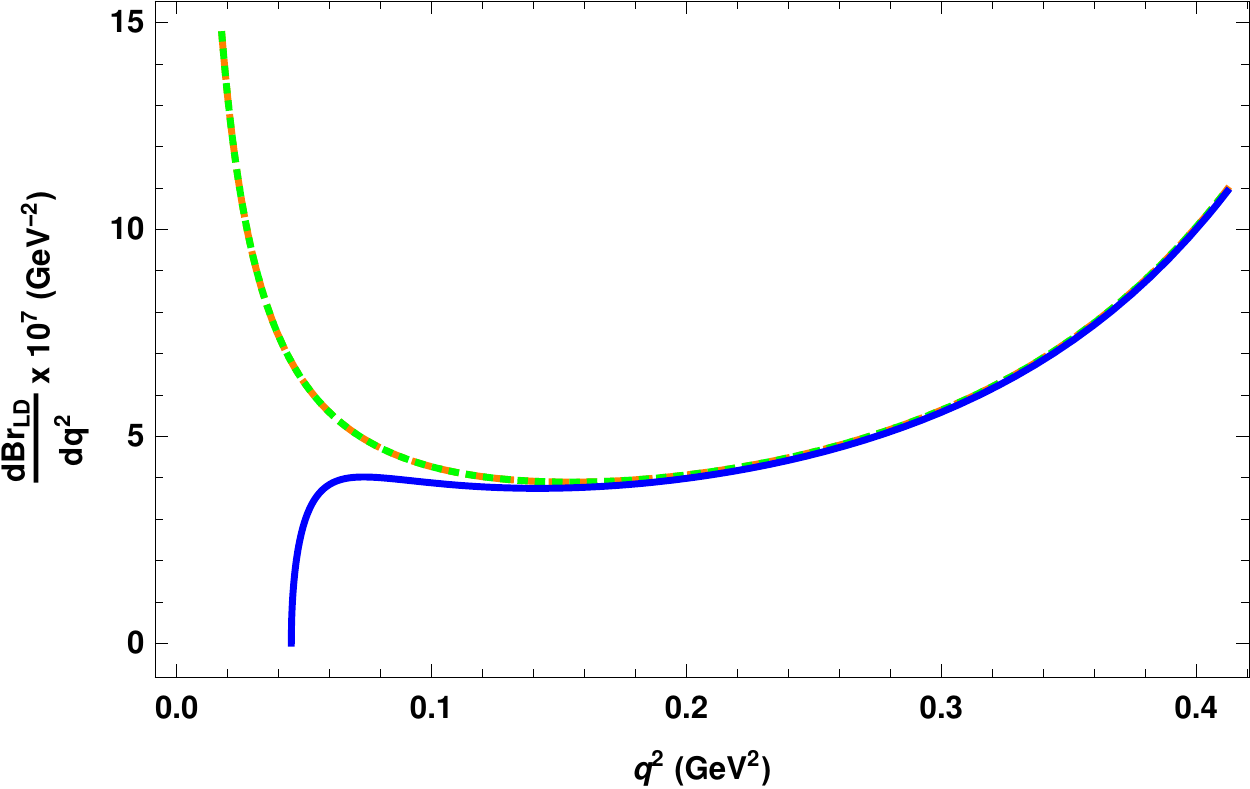}
    \caption{$B\to K^\ast \ell^+ \ell^-$ long distance differential branching ratios as a function of the dilepton invariant mass squared. Units are GeV$^{-2}$, while colors refer to $m_\ell=0$ case in orange (dashed), $m_\ell=m_e$ in green (dotted) and $m_\ell=m_\mu$ in blue (solid).}
    \label{f:dBrdq2LD}
\end{figure}

\noindent
The last ingredient of the calculation is the $\omega \to \pi \gamma$ transition. The radiative decays of the light vector and axial-vector mesons have been systematically studied in Ref.  \cite{Lutz:2008km} in the framework of chiral Lagrangian  written in terms of the Goldstone
boson octet and the nonet of light vector mesons. The electromagnetic form factor relevant to the $\omega \to \pi \ell^+\ell^-$ 
has been obtained in Ref. \cite{Terschlusen:2010gtc}, where the case with $\ell\equiv e , \mu$ has been considered. It is worth noting 
that the corresponding widths differ of about one order of magnitude due to the deep decreasing form factor at small dilepton invariant mass.
In fact, experimentally, we have $Br(\omega\to\pi^0 e^+e^-)= (7.7\pm0.6)\times 10^{-4}$ and 
$Br(\omega\to\pi^0 \mu^+\mu^-)= (1.34\pm0.18)\times 10^{-4}$ \cite{Workman:2022ynf}.\\ Schematically, the LD amplitude can be written as
\begin{eqnarray}
&&A_{LD}(\lambda_{K^\ast},\sigma_{\ell^+},\sigma_{\ell^-}) ~=~  
\frac{1}{q^2} \sum_{\lambda_\omega,\lambda_\gamma} \mathcal{A}(B\to K \omega(\lambda_\omega))       \times\nonumber\\
&& 
 \hspace{1.2truecm} \mathcal{A}(K \omega(\lambda_\omega)\to K^\ast(\lambda_{K^\ast})  \gamma^\ast(\lambda_\gamma))~ \mathcal{A}(\gamma^\ast(\lambda_\gamma) \to \ell^+(\sigma_{\ell^+}) \ell^-(\sigma_{\ell^-})),
\label{e:ALD1}
\end{eqnarray}
where $\lambda$'s and $\sigma$'s are the vector particle and fermion polarizations, respectively. Whereas, the weak decay of B into $K \omega$ and the rescattering amplitude can be recast as follows: 
\begin{eqnarray}
&& \mathcal{A}(B\to K \omega(\lambda_\omega)) = g_{BK\omega}~ (p_K + p_\omega)\cdot \epsilon^\ast(\lambda_\omega), \\
&& \mathcal{A}(K \omega(\lambda_\omega)\to K^\ast(\lambda_{K^\ast})  \gamma^\ast(\lambda_\gamma)) =
\frac{1}{4\pi} \int_0^{2\pi} d\phi~ \int_{t_{min}}^{t_{max}} \frac{dt}{2 |\vec p_\omega| |\vec q|}\times\nonumber \\
&&\hspace{2.0truecm}\left\{  \frac{i\, e~g_{KK^\ast \pi}~f_{\omega\pi^0}}{t - m_\pi^2}\, p_K\cdot \epsilon^\ast(\lambda_K) \epsilon^{\mu\nu\alpha\beta}p_{\omega\mu}q_\nu\epsilon_\alpha(\lambda_\omega)\epsilon_\beta^\ast(\lambda_\gamma)\right\},\\
&& \mathcal{A}(\gamma^\ast(\lambda_{\gamma}) \to \ell^+(\sigma_{\ell^+}) \ell^-(\sigma_{\ell^-})) = e~ \overline u_{\ell^-}(\sigma_{\ell^-}) \left( \gamma \cdot \epsilon_\gamma(\lambda_\gamma) \right) v_{\ell^+}(\sigma_{\ell^+}),
\end{eqnarray}
being $\phi$ the euclidean $\vec p_\omega$ azimuth (z-axis) and ${\displaystyle t = (p_{K^\ast}-p_K)^2 = (p_\omega - q)^2}$\footnote{t$_{min}$ (t$_{max}$) corresponds to 
t when the euclidean $\vec p_\omega$ colatitude $\theta$ equals 0 ($\pi$).}. Here $f_{\omega\pi^0}$ is the electromagnetic form factor computed in \cite{Terschlusen:2010gtc}, $g_{BK\omega}$ contains the weak coupling and the CKM matrix elements, $u$ and $v$ are Dirac spinors. Our knowledge of the $f_{\omega\pi^0}$ electromagnetic form factor is dictated by the physical range of the $\omega \to \pi^0\gamma^*$ transition, $\displaystyle [4 m_\ell^2,(m_\omega-m_\pi)^2]$, and, accordingly, our results are valid in the same range of the diletpton invariant mass. The raising behavior of the branching ratios is due to the $\rho$ meson pole in the $f_{\omega\pi^0}$ electromagnetic form factor. Due to the pseudoscalar nature of the $B$ meson, only the longitudinal polarization of the $\omega$ meson contributes to the amplitude; furthermore, after integrating on the azimuth angle ($\phi$), the positive (negative) $K^\ast$ meson polarization selectively couples to the negative (positive) $\gamma^\ast$ polarization 
(while the $\gamma^\ast$ longitudinal polarization is ruled out by the $\omega$-$\gamma^\ast$-$\pi$ Levi--Civita coupling). In Fig. \ref{f:dBrdq2LD} the differential branching ratio $dBr(B\to K^\ast \ell^+ \ell^-)_{LD}/dq^2$
(in unit of $10^{-7}$) is plotted vs $q^2$. The long distance part of the branching ratio is, in the three different cases $m_\ell=(0,m_e, m_\mu)$ (with colors orange, green and blue, respectively), of the same order of magnitude of the short distance part as we shall see later. In this range of $q^2$ the difference between the electron and the muon case in the $\omega \to \pi^0 \ell^+ \ell^-$ leads to a small lepton flavour violation. In fact, the branching ratios evaluated in this range of $q^2$ give $Br(B\to K^\ast e^+ e^-)_{LD} = 2.0\times 10^{-7}$ and $Br(B\to K^\ast \ell^+ \ell^-)_{LD} = 1.9\times 10^{-7}$ this effect which is not related to any new interaction violating the lepton flavor universality fakes the violation.\par
\begin{figure}[!htp]
    \centering
    \includegraphics[width=0.8\textwidth]{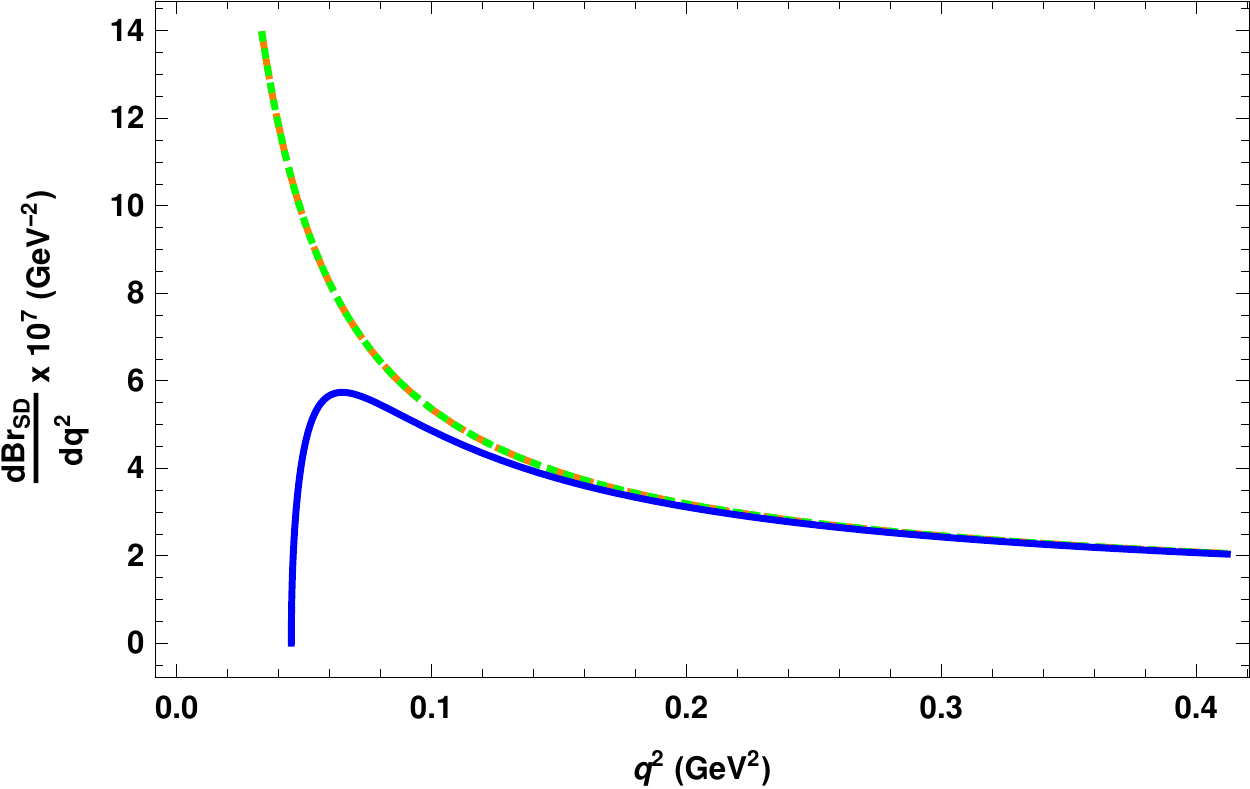}
    \caption{$B\to K^\ast \ell^+ \ell^-$ short distance differential branching ratios as a function of the dilepton invariant mass squared. Units are GeV$^{-2}$, while colors refer to $m_\ell=0$ case in orange (dashed), $m_\ell=m_e$ in green (dotted) and $m_\ell=m_\mu$ in blue (solid).}
    \label{f:dBrdq2SD}
\end{figure}

 In order to understand to what extent the long distance contribution to $B\to K^\ast \ell^+\ell^-$ can affect both the branching ratio and the lepton flavour universality violation, it is necessary to compute the short distance amplitude, i.e. the amplitude in Eq. (\ref{e:SD}). We employ the set of form factors calculated in Ref \cite{Colangelo:1995jv} and evaluate the $dBr(B\to K^\ast \ell^+\ell^-)/dq^2$ with lepton finite mass in the final state. In Fig. \ref{f:dBrdq2SD} we plot the short distance differential branching ratio alone for the case of $m_\ell=0$ in orange, $m_\ell= m_e$ in green and $m_\ell= m_\mu$ in blue; it is worth noting that the first two curves overlap exactly (although not analytically).


\begin{figure}[h]
    \centering
    \includegraphics[width=0.8\textwidth]{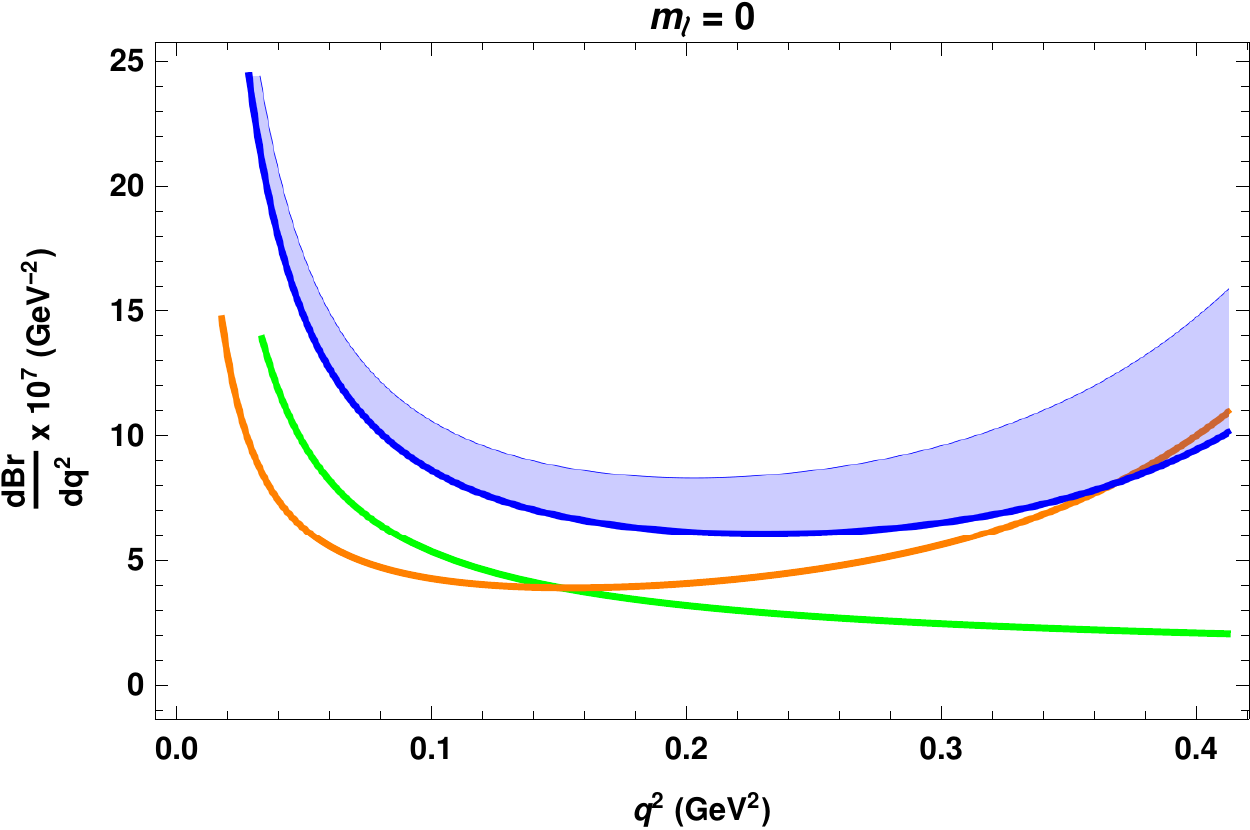}
    \caption{$B\to K^\ast \ell^+ \ell^-$ total differential branching ratios as a function of the dilepton invariant mass squared when $m_\ell=0$ (units are GeV$^{-2}$). Colors refer to the short distance contribution in green, the long distance contribution in orange, while the combination of both contributions spans an area in blue (upper thin bound corresponds to 
    the difference, lower thick one to the sum).}
    \label{f:dBrdq2SDpmLD_0}
\end{figure}
\begin{figure}[h]
    \centering
    \includegraphics[width=0.8\textwidth]{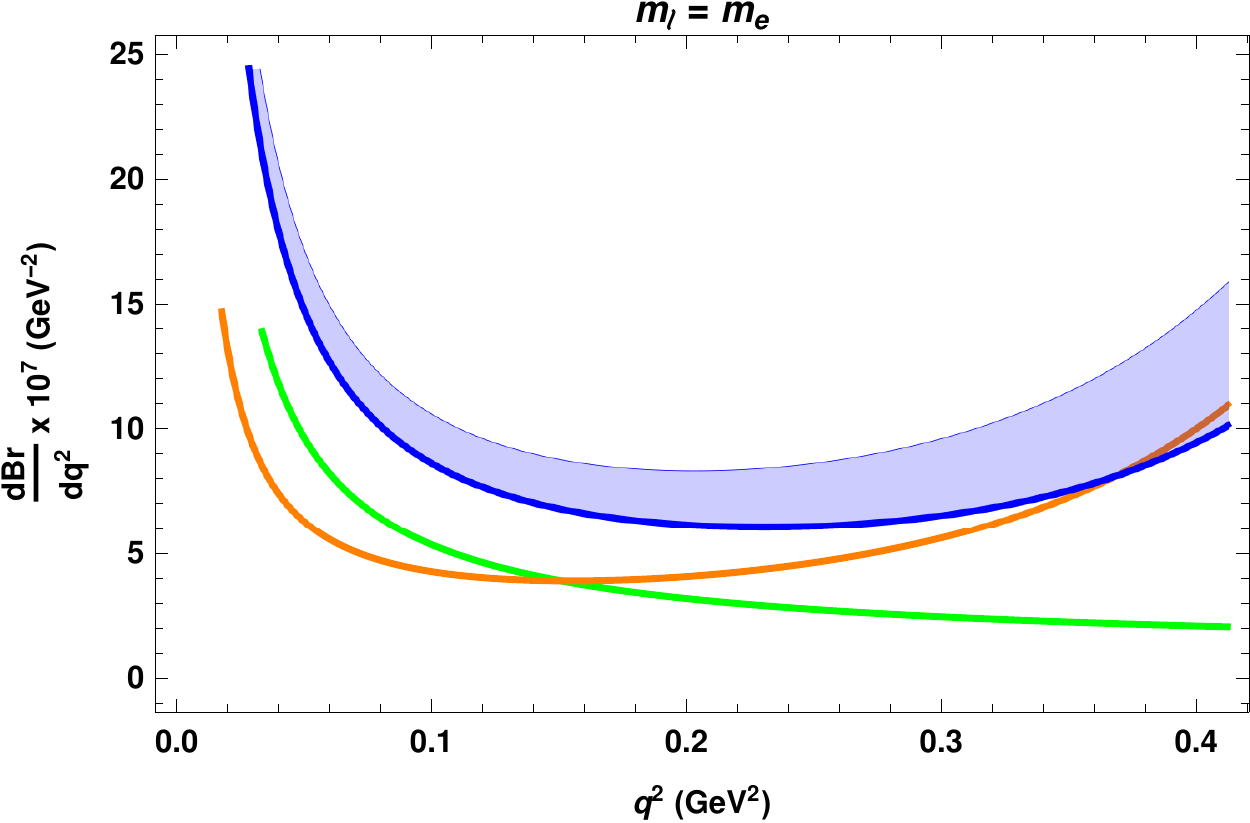}
    \caption{$B\to K^\ast \ell^+ \ell^-$ total differential branching ratios as a function of the dilepton invariant mass squared when $m_\ell=m_e$ (units are GeV$^{-2}$). Colors refer to the short distance contribution in green, the long distance contribution in orange, while the combination of both contributions spans an area in blue (upper thin bound corresponds to 
    the difference, lower thick one to the sum).}
    \label{f:dBrdq2SDpmLD_e}
\end{figure}
\begin{figure}[h]
    \centering
    \includegraphics[width=0.8\textwidth]{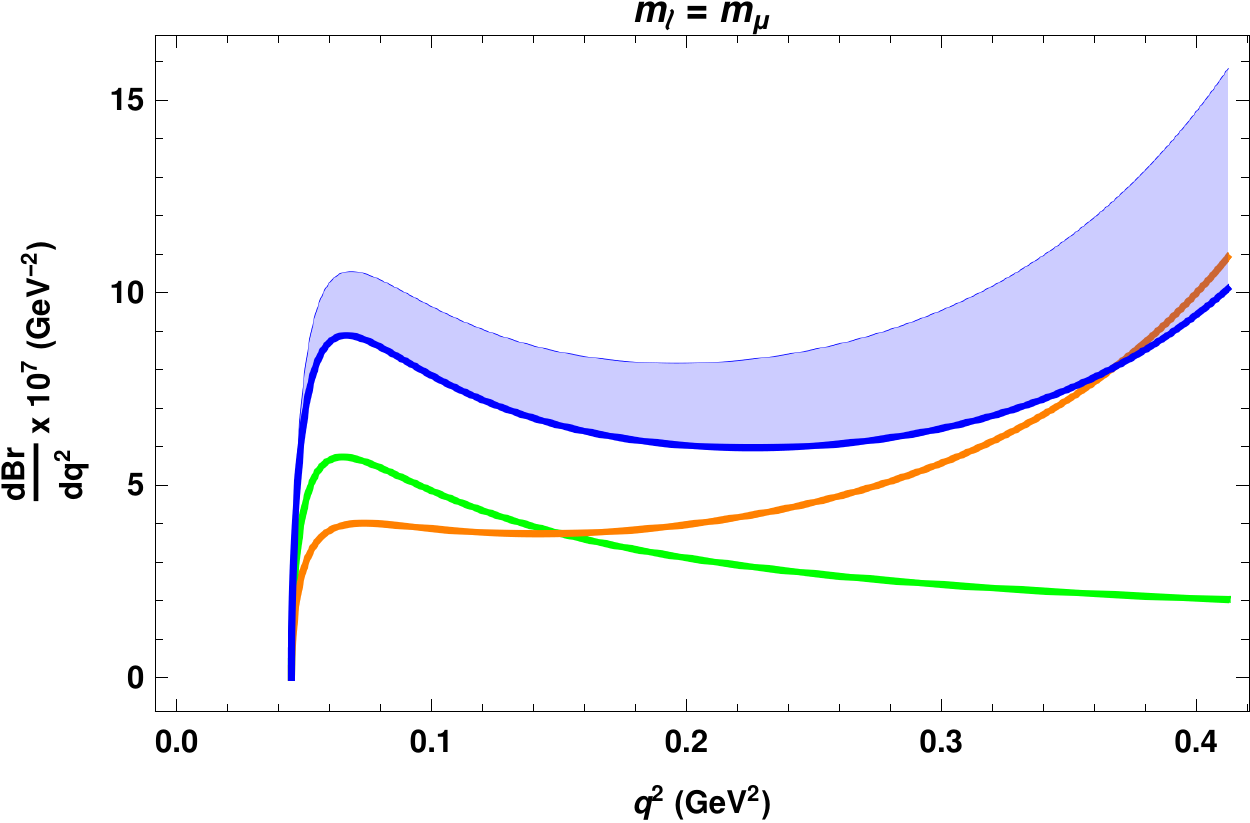}
    \caption{$B\to K^\ast \ell^+ \ell^-$ total differential branching ratios as a function of the dilepton invariant mass squared when $m_\ell=m_\mu$ (units are GeV$^{-2}$). Colors refer to the short distance contribution in green, the long distance contribution in orange, while the combination of both contributions spans an area in blue (upper thin bound corresponds to 
    the difference, lower thick one to the sum).}
    \label{f:dBrdq2SDpmLD_mu}
\end{figure}
The figures \ref{f:dBrdq2SDpmLD_0},
\ref{f:dBrdq2SDpmLD_e},
\ref{f:dBrdq2SDpmLD_mu} show that the long distance contribution (in orange) is of the same order of magnitude of the short distance one regardless of the lepton flavor. The long distance contribution in all cases increases approaching the $(m_\omega-m_\pi)^2$ upper limit for $q^2$ because of the pole dominance of the form factor $f_{\omega\pi^0}$. The blue regions represent the band of values of the total differential branching ratios: the lower (upper) curves refer to the constructive (distructive) interference between the short and the long distance contributions. For each $q^2$ the width of the band can be considered as an estimation of the theoretical error on the differential branching ratio calculation. 
In table \ref{t:BrRKs} the {\it partial} branching ratios and $R_{K^\ast}$, i.e. evaluated in the limited dilepton mass squared range $\displaystyle [4 m_\mu^2,(m_\omega-m_\pi)^2]$, are collected to point out the amount of the 
long distance contribution: it is clear that, in the region of $q^2$ studied, the long distance contribution increase the tension with the experimental data on $R_{K^\ast}$. In fact, the measurement of $R_{K^\ast}$ in the smallest range 
was done by LHCb collaboration ~\cite{LHCb:2017avl} 
	\be
	R_{K^*}  = 
		0.660^{+0.110}_{-0.070} \pm 0.024 \hspace{1cm} 
		(2m_\mu)^2 < q^2 < 1.1\, {\rm GeV^2}  \ .
	\ee
Moreover, the {\sc Belle} collaboration presented the following preliminary result~\cite{Belle:2019oag} obtained by averaging over $B^0$ and $B^+$: 
\be
R_{K^*}[0.045,1.1]=0.52^{+0.36}_{-0.26} \pm 0.05.
\ee
These values have to be compared to the SM predictions \cite{Bordone:2016gaq}
\be
R^{SM}_{K^*}  =
		0.906 \pm 0.028 \hspace{1cm}  (2m_\mu)^2 < q^2 < 1.1  {\rm GeV^2}\ ,  
\ee
which is consistent with our value of $R_{K^\ast}$ by considering SD contribution alone. The inclusion of the LD term increases our prediction of $R_{K^\ast}$.
\par
Before concluding, it is essential to emphasize that the contribution we have calculated cannot be rewritten in terms of the SD amplitude because it is essentially a rescattering and so the Lorentz couplings are different from the ones in Eq. (\ref{e:SD}). This entails that it cannot be estimated by fitting data to the shifts of the Wilson coefficients $C_9$ and $C_{10}$ with respect to the Standard Model values.

In conclusion, in this letter we have estimated the LD contribution of the light quark loop to the $B\to K^\ast \ell^+\ell^-$ with $m_\ell\in\{m_e,m_\mu\}$. We have focused on a specific hadronic rescattering channel: $B\to K\omega\to K^\ast\ell^+\ell^-$. The calculations have been performed in the range of the dilepton mass squared $\left [4 m_\mu^2,(m_\omega-m_\pi)^2\right]$, where our 
hadronic representation of the quark loop is reliable. The LD contribution increases the branching ratios of about a factor 2.5 with respect to the SD results. Our findings also indicate a change in the ratio $R_{K^\ast}$.

\begin{table}[htp]
\centering
\begin{tabular}{|ccccc|}
\hline
    & \textbf{SD} & \textbf{LD} & \textbf{SD+LD}   &  \textbf{SD-LD} \\ 
\hline
$m_\ell=0, m_e~(\times 10^{-7})$    & 1.333 & 2.028	& 2.837 & 3.885 \\
$m_\ell=m_\mu~(\times 10^{-7})$ 	& 1.190 & 1.919	& 2.609 & 3.609 \\
R$_{K^\ast}$                    	& 0.893 & 0.946	& 0.920 & 0.929 \\
\hline
\end{tabular}
\caption{Branching ratios and R$_{K^\ast}$ computed in the dilepton mass
squared range 
$\displaystyle [4 m_\mu^2,(m_\omega-m_\pi)^2]$ as a function of lepton 
mass ($m_\ell$) 
and along with selected amplitude contributions (\textbf{SD},\textbf{LD} and
combinations thereof).}
\label{t:BrRKs}
\end{table}
 
\section*{Acknowledgments}
\noindent
We thank Stefan Leupold for discussions.


\begin{thebibliography}{99}

\bibitem{DAlise:2022ypp}
A.~D'Alise, G.~De Nardo, M.~G.~Di Luca, G.~Fabiano, D.~Frattulillo, G.~Gaudino, D.~Iacobacci, M.~Merola, F.~Sannino and P.~Santorelli, \textit{et al.}
[arXiv:2204.03686 [hep-ph]].

\bibitem{GSW}
B. Grinstein, M.J. Savage and M. B. Wise, Nucl. Phys. B {\bf 319}, 271 (1989).

\bibitem{Chetyrkin:1996vx}
K.~G.~Chetyrkin, M.~Misiak and M.~Munz,
Phys. Lett. B \textbf{400}, 206-219 (1997)
[erratum: Phys. Lett. B \textbf{425}, 414 (1998)]
doi:10.1016/S0370-2693(97)00324-9
[arXiv:hep-ph/9612313 [hep-ph]].


\bibitem{noibis}
M. Ladisa and P. Santorelli, in preparation.

\bibitem{Colangelo:1995jv}
P.~Colangelo, F.~De Fazio, P.~Santorelli and E.~Scrimieri,
Phys. Rev. D \textbf{53}, 3672-3686 (1996)
[erratum: Phys. Rev. D \textbf{57}, 3186 (1998)]
doi:10.1103/PhysRevD.53.3672
[arXiv:hep-ph/9510403 [hep-ph]].


\bibitem{Gubernari:2020eft}
N.~Gubernari, D.~van Dyk and J.~Virto,\\
JHEP \textbf{02}, 088 (2021)
doi:10.1007/JHEP02(2021)088
[arXiv:2011.09813 [hep-ph]].


\bibitem{Beneke:2001at}
M.~Beneke, T.~Feldmann and D.~Seidel,
Nucl. Phys. B \textbf{612}, 25-58 (2001)
doi:10.1016/S0550-3213(01)00366-2
[arXiv:hep-ph/0106067 [hep-ph]].



\bibitem{Khodjamirian:2010vf}
A.~Khodjamirian, T.~Mannel, A.~A.~Pivovarov and Y.~M.~Wang,
JHEP \textbf{09}, 089 (2010)
doi:10.1007/JHEP09(2010)089
[arXiv:1006.4945 [hep-ph]].

\bibitem{Gubernari:2022hxn}
N.~Gubernari, M.~Reboud, D.~van Dyk and J.~Virto,
[arXiv:2206.03797 [hep-ph]].

\bibitem{Ciuchini:2021smi}
M.~Ciuchini, M.~Fedele, E.~Franco, A.~Paul, L.~Silvestrini and M.~Valli,
[arXiv:2110.10126 [hep-ph]].


\bibitem{Isola:2003fh}
C.~Isola, M.~Ladisa, G.~Nardulli and P.~Santorelli,
Phys. Rev. D \textbf{68}, 114001 (2003)
doi:10.1103/PhysRevD.68.114001
[arXiv:hep-ph/0307367 [hep-ph]].


\bibitem{Workman:2022ynf}
R.~L.~Workman \textit{et al.} [Particle Data Group],
PTEP \textbf{2022}, 083C01 (2022)
doi:10.1093/ptep/ptac097


\bibitem{Ciuchini:1997hb}
M.~Ciuchini, E.~Franco, G.~Martinelli and L.~Silvestrini,
Nucl. Phys. B \textbf{501}, 271-296 (1997)
doi:10.1016/S0550-3213(97)00388-X
[arXiv:hep-ph/9703353 [hep-ph]].

\bibitem{Isola:2001ar}
C.~Isola, M.~Ladisa, G.~Nardulli, T.~N.~Pham and P.~Santorelli,
Phys. Rev. D \textbf{64}, 014029 (2001)
doi:10.1103/PhysRevD.64.014029
[arXiv:hep-ph/0101118 [hep-ph]].

\bibitem{Isola:2001bn}
C.~Isola, M.~Ladisa, G.~Nardulli, T.~N.~Pham and P.~Santorelli,
Phys. Rev. D \textbf{65}, 094005 (2002)
doi:10.1103/PhysRevD.65.094005
[arXiv:hep-ph/0110411 [hep-ph]].


\bibitem{Ladisa:2004bp}
M.~Ladisa, V.~Laporta, G.~Nardulli and P.~Santorelli,
Phys. Rev. D \textbf{70}, 114025 (2004)
doi:10.1103/PhysRevD.70.114025
[arXiv:hep-ph/0409286 [hep-ph]].

\bibitem{Deandrea:2005ii}
A.~Deandrea, M.~Ladisa, V.~Laporta, G.~Nardulli and P.~Santorelli,
Int. J. Mod. Phys. A \textbf{21}, 4425-4438 (2006)
doi:10.1142/S0217751X0603401X
[arXiv:hep-ph/0508083 [hep-ph]].



\bibitem{Lutz:2008km}
M.~F.~M.~Lutz and S.~Leupold,
Nucl. Phys. A \textbf{813}, 96-170 (2008)\\
doi:10.1016/j.nuclphysa.2008.09.005
[arXiv:0801.3821 [nucl-th]].


\bibitem{Terschlusen:2010gtc}
C.~Terschlusen and S.~Leupold,
Phys. Lett. B \textbf{691}, 191-201 (2010)\\
doi:10.1016/j.physletb.2010.06.033
[arXiv:1003.1030 [hep-ph]].



\bibitem{LHCb:2017avl}
R.~Aaij \textit{et al.} [LHCb],
JHEP \textbf{08} (2017), 055
doi:10.1007/JHEP08(2017)055
[arXiv:1705.05802 [hep-ex]].
%
%
\bibitem{Belle:2019oag}
A.~Abdesselam \textit{et al.} [Belle],
Phys. Rev. Lett. \textbf{126} (2021) no.16, 161801
doi:10.1103/PhysRevLett.126.161801
[arXiv:1904.02440 [hep-ex]].
%
%
\bibitem{Bordone:2016gaq}
M.~Bordone, G.~Isidori and A.~Pattori,
Eur. Phys. J. C \textbf{76} (2016) no.8, 440
doi:10.1140/epjc/s10052-016-4274-7
[arXiv:1605.07633 [hep-ph]].
	

			
\end{thebibliography}
\end{document}